\pdfoutput=1

\documentclass[11pt,twoside,a4paper,cmspaper,final,collab]{cms-tdr}
\def\svnVersion{215791}\def\cmsCernNo{2013-157}\def\cmsCernDate{\today}\def\cmsMessage{Published in Physics Letters B as \href{http://dx.doi.org/10.1016/j.physletb.2013.10.016}{\doi{10.1016/j.physletb.2013.10.016.}}}

\begin{document}\cmsNoteHeader{BPH-11-016}

\hyphenation{had-ron-i-za-tion}
\hyphenation{cal-or-i-me-ter}
\hyphenation{de-vices}

\RCS$Revision: 209977 $
\RCS$HeadURL: svn+ssh://svn.cern.ch/reps/tdr2/papers/BPH-11-016/trunk/BPH-11-016.tex $
\RCS$Id: BPH-11-016.tex 209977 2013-10-03 18:35:35Z kfjack $
\newlength\cmsFigWidth
\ifthenelse{\boolean{cms@external}}{\setlength\cmsFigWidth{0.95\columnwidth}}{\setlength\cmsFigWidth{0.80\textwidth}}
\ifthenelse{\boolean{cms@external}}{\providecommand{\cmsLeft}{top}}{\providecommand{\cmsLeft}{left}}
\ifthenelse{\boolean{cms@external}}{\providecommand{\cmsRight}{bottom}}{\providecommand{\cmsRight}{right}}
\providecommand{\cPXb}{\ensuremath{\cmsSymbolFace{X}_\cmsSymbolFace{b}}\xspace} 
\providecommand{\cPXn}{\ensuremath{\cmsSymbolFace{X}}(3872)\xspace} 
\cmsNoteHeader{BPH-11-016} 
\title{Search for a new bottomonium state decaying to $\PgUa\Pgpp\Pgpm$ in pp collisions at $\sqrt{s} = 8$\TeV}

\date{\today}

\abstract{
The results of a search for the bottomonium counterpart, denoted as \cPXb, of the exotic charmonium state $\cPXn$ is presented.
The analysis is based on a sample of pp collisions at $\sqrt{s} = 8$\TeV collected by the CMS experiment at the LHC,
corresponding to an integrated luminosity of 20.7\fbinv.
The search looks for
the exclusive decay channel $\cPXb \to \PgUa\Pgpp\Pgpm$ followed by $\PgUa \to \Pgmp\Pgmm$.
No evidence for an \cPXb signal is observed.
Upper limits are set at the 95\% confidence level on the ratio of the
inclusive production cross sections times the branching fractions to
$\PgUa\Pgpp\Pgpm$ of the $\cPXb$ and the $\PgUb$.  The upper limits on the ratio
are in the range 0.9--5.4\% for $\cPXb$ masses between 10 and 11\GeV.
These are the first upper limits on the production of a possible $\cPXb$ at
a hadron collider.
}

\hypersetup{%
    pdfauthor={CMS Collaboration},%
    pdftitle={Search for a new bottomonium state decaying to Upsilon(1S) pi+ pi- in pp collisions at sqrt(s) = 8 TeV},%
    pdfsubject={CMS},%
    pdfkeywords={CMS, physics, exotic quarkonia}
}

\maketitle 

\section{Introduction}

In the past decade, several unexpected charmonium states, such as the $\cPXn$ and the Y(4260), have been discovered~\cite{Choi:2003ue,Aubert:2005rm} and then confirmed~\cite{Aubert:2004ns, Yuan:2007sj}
by the Belle and BaBar experiments.
The $\cPXn$ state has also been seen by hadron collider experiments~\cite{Acosta:2003zx, Abazov:2004kp, Aaij:2011sn, Chatrchyan:2013cld}.
These exotic states are not predicted in well-established theoretical models,
and progress in this area has been mostly driven by experiments.
Interpretations as tetraquark states or hadronic molecules have been proposed~\cite{AlFiky:2005jd,Ebert:2005nc,Ebert:2008se,Matheus:2006xi}.
It is therefore natural to look for
similar states in the bottomonium system~\cite{Hou:2006it,AlFiky:2005jd,Ebert:2005nc,Ebert:2008se,Matheus:2006xi,Guo:2013sya}.
Finding such exotic bottomonium states may provide a more complete picture of exotic quarkonia and help to clarify their production mechanisms and intrinsic properties.

The exotic resonance $\cPXn$ was discovered in the final state $\JPsi \Pgpp\Pgpm$.
The CDF~\cite{Acosta:2003zx} and CMS~\cite{Chatrchyan:2013cld} experiments have
shown that the $\cPXn$ is produced not only
through \PB-meson decays, but also through prompt production.
At the Large Hadron Collider (LHC),
CMS has measured the prompt production of the $\cPXn$ to be about 80\% of the total production~\cite{Chatrchyan:2013cld}.
A bottomonium counterpart of the $\cPXn$, denoted as $\cPXb$, would be expected to decay through $\cPXb \to \PgUa\Pgpp\Pgpm$. Several known properties of the $\cPXn$ state provide clues in the search for the $\cPXb$.
The $\cPXn$ has a small natural width $<1.2$\MeV~\cite{Beringer:1900zz} and its production rate times the ${\cPXn} \to \JPsi \Pgpp\Pgpm$
branching fraction
is $(6.56\pm0.29\pm0.65)$\% of the corresponding $\psi(2S) \to \JPsi \Pgpp\Pgpm$ value~\cite{Chatrchyan:2013cld}, roughly independent of transverse momentum (\pt).
In analogy, the $\cPXb$ state could also be a narrow resonance, with a sizable decay
rate in the $\PgUa\Pgpp\Pgpm$ final state.
The mass of the $\cPXb$
is predicted in some models~\cite{Hou:2006it,AlFiky:2005jd} to be close to either the $\PB\PaB$ or $\PB\PaB^{*}$
thresholds, 10.562 and 10.604\GeV, respectively. Alternative models~\cite{Ebert:2005nc,Ebert:2008se,Matheus:2006xi} suggest
other possibilities in the 10--11\GeV range.

A search in the $\PgUa\Pgpp\Pgpm$ final state has been performed by the Belle experiment in the mass region
around the $\PgU\mathrm{(10860)}$ resonance~\cite{Abe:2007tk};
a statistically significant excess of events was found but its origin remains inconclusive~\cite{Adachi:2008pu}. The BaBar experiment found an $\Upsilon\mathrm{(1D)}$
bottomonium resonance in the same final state at 10.165\GeV~\cite{delAmoSanchez:2010kz}.
However, the production rates for these two states are expected to be very small at the LHC~\cite{Ali:2011qi, Brambilla:2010cs}, and they are not expected to be seen in the present analysis.

This Letter presents the results of a search
for the exotic bottomonium state $\cPXb$ through its decay $\cPXb \to \PgUa\Pgpp\Pgpm$,
using a data sample of $\Pp\Pp$ collisions at $\sqrt{s} = 8$\TeV collected by the CMS experiment at the LHC and corresponding to an integrated luminosity of 20.7\fbinv.

The strategy of this analysis is to search for a peak, other than the
known $\PgUb$ and $\PgUc$ resonances, in the $\PgUa(\Pgmp\Pgmm)\Pgpp\Pgpm$ invariant-mass spectrum.
The results are presented in terms of
the relative inclusive production cross sections of the
$\cPXb$ and $\PgUb$ states times their decay branching fractions to $\PgUa\Pgpp\Pgpm$,
$\sigma(\Pp\Pp \to \cPXb \to \PgUa\Pgpp\Pgpm)/\sigma(\Pp\Pp\to \PgUb \to \PgUa\Pgpp\Pgpm)$,
as a function of the $\cPXb$ mass between 10 and 11\GeV.
Many systematic uncertainties cancel in this ratio.
The analysis probes the kinematic region $\pt(\PgUa\Pgpp\Pgpm)>13.5$\GeV and $\abs{y(\PgUa\Pgpp\Pgpm)}<2.0$, where $y$ denotes the rapidity.

\section{CMS detector and trigger}

The central feature of the CMS apparatus is a superconducting solenoid of 6\unit{m} internal diameter, providing a magnetic field of 3.8\unit{T}. Within the magnet volume are a silicon pixel and strip tracker, a lead tungstate crystal electromagnetic calorimeter, and a brass/scintillator hadron calorimeter.
The inner tracker measures charged particles within the pseudorapidity range $\abs{\eta}< 2.5$, where
$\eta = -\ln[\tan(\theta/2)]$ and $\theta$ is the polar angle with respect to the counterclockwise proton beam direction.
Muons are measured with detection planes made using three technologies: drift tubes, cathode strip chambers, and resistive-plate chambers. Matching muons to tracks measured in the silicon tracker results in a \pt resolution
better than 1.5\% for a typical muon in this analysis.
Extensive forward calorimetry complements the angular coverage provided by the barrel and endcap detectors.
A more detailed description of the CMS detector can be found in Ref.~\cite{:2008zzk}.

A two-level trigger system~\cite{Adam:2005zf} selects events for further analysis.
The events analyzed in this search are collected with a trigger
requiring at least two oppositely charged muon candidates.  The
invariant mass of the muon pair is required to be in the range between 8.5 and 11.5\GeV.
The transverse momentum of the muon pair must be above 6.9\GeV,
and the dimuon vertex fit $\chi^2$ probability must be greater than 0.5\%.

\section{Reconstruction and event selection}

The reconstruction of the $\cPXb$ and $\PgUb$ candidates
begins with the reconstruction of an $\PgUa \to \Pgmp\Pgmm$ candidate and two additional charged tracks in the event, assuming the pion mass hypothesis for the latter.
The  $\PgUa$ candidate is reconstructed as a pair of oppositely charged muons.
Each muon is required to leave hits in at least six tracker
layers, at least two of which must be in the silicon pixel detector,
and to be matched with at least one segment in the muon system.
Muons are required to have $\pt > 2.5$\GeV, $\abs{\eta}<2.1$, and good track-fit quality.
The muon tracks are required to intersect the beam line within a cylinder of 3\unit{cm} in radius and 30\unit{cm} in length
around the primary vertex position, which is selected as the vertex with the largest sum of $\pt^2$ of the
tracks associated with it.
The rapidity of the muon pair is required to be within $\abs{y(\Pgmp\Pgmm)} < 2.0$.
Each $\PgUa$ candidate must have $\pt > 13.5$\GeV and the
$\chi^2$ probability from the dimuon vertex fit is required to be larger than 1\%;
these requirements are more stringent than those used by the trigger.
The $\PgUa$ invariant-mass window is defined as the region within
$\pm 2.5\sigma_M$ of the $\PgUa$ mass~\cite{Beringer:1900zz}, where the dimuon
mass resolution, $\sigma_M$, is measured to increase from 65 to 125\MeV as $\abs{y(\Pgmp\Pgmm)}$ changes from 0 to 2.

The $\PgUa\Pgpp\Pgpm$ candidate is reconstructed
by combining two oppositely charged tracks with the $\PgUa$ candidate.
The pions must each have a minimum \pt of 400\MeV, be in
the region $\abs{\eta}<2.5$, have a track fit $\chi^2/\mathrm{ndf} < 5$, and at least
11 strip tracker hits and two silicon pixel hits.
In order to reduce the combinatorial background from
additional tracks in the event, the $\chi^2$
probability of the dipion vertex fit is required to be
larger than 10\% and the pions are required to be within a radius
$\Delta R = \sqrt{(\Delta\eta)^2 + (\Delta\phi)^2}<0.7$
with respect to the $\PgUa$ direction,
where $\Delta\eta$ ($\Delta\phi$) is
the difference in pseudorapidity (azimuthal angle in radians) between the pion direction and the $\PgUa$ momentum.
A common-vertex constrained fit is applied to the tracks forming the $\PgUa(\Pgmp\Pgmm)\Pgpp\Pgpm$ candidate, together with the constraint that the dimuon invariant mass be equal to the world-average $\PgUa$ mass~\cite{Beringer:1900zz}.
The $\chi^2$ probability from the vertex-mass constrained fit is required
to be greater than 10\%.
There are on average  2.3 reconstructed candidates per event for the events with at least one candidate; all of the reconstructed candidates are included in the analysis.
From simulations the combinatorial background candidates are found to be randomly distributed in mass and
do not introduce any bias in the extraction of the signal.
The selection criteria described above
were optimized using
a genetic algorithm~\cite{Hocker:2007ht} that maximized the
expected significance of the signal in the mass region near the $\PgUb$.
The statistical significance of the signal is expected to be more than five standard deviations
if the $\cPXb$ branching fraction times production cross section relative to the $\PgUb$ is 6.56\% (or higher), analogous to that
of the $\cPXn$ relative to the $\psi(2S)$.

The simulated samples of the $\PgUb$ and $\cPXb$ events are generated with the \PYTHIA 6.426~\cite{Sjostrand:2006za} event generator, and the particle decays are
modeled using the \EVTGEN package~\cite{Lange:2001uf},
assuming that the $\PgUb$ and $\cPXb$ states have the same production mechanism and are both produced unpolarized.
The unpolarized assumption for the $\PgUb$ is supported by a recent CMS measurement~\cite{Chatrchyan:2012woa}.
 In the event generation, the
$\cPXb$ state is assumed to be a narrow resonance with the same quantum numbers as the $\PgUb$.
Generated events are processed through a full detector simulation based on \GEANTfour~\cite{Agostinelli:2002hh}.
The simulated $\PgUb$ events are
reweighted according to the dipion invariant-mass spectrum in $\PgUb \to \PgUa\Pgpp\Pgpm$ decays measured by the CLEO experiment~\cite{CroninHennessy:2007zz}.
The kinematic distributions in the simulated $\PgUb$ sample are compared with the corresponding
distributions in the data and found to be in good agreement.
The $\cPXb$ events are assumed to have the same dipion mass distribution as the $\PgUb$ events, which is very similar to the dipion mass distribution of the $\cPXn$~\cite{Chatrchyan:2013cld}.
Other possible $\cPXb$ decay models are considered as systematic uncertainties, as described in
Sec.~\ref{sec:systematics}.

\begin{figure}[thb]
  \centering
    \includegraphics[width=0.49\textwidth]{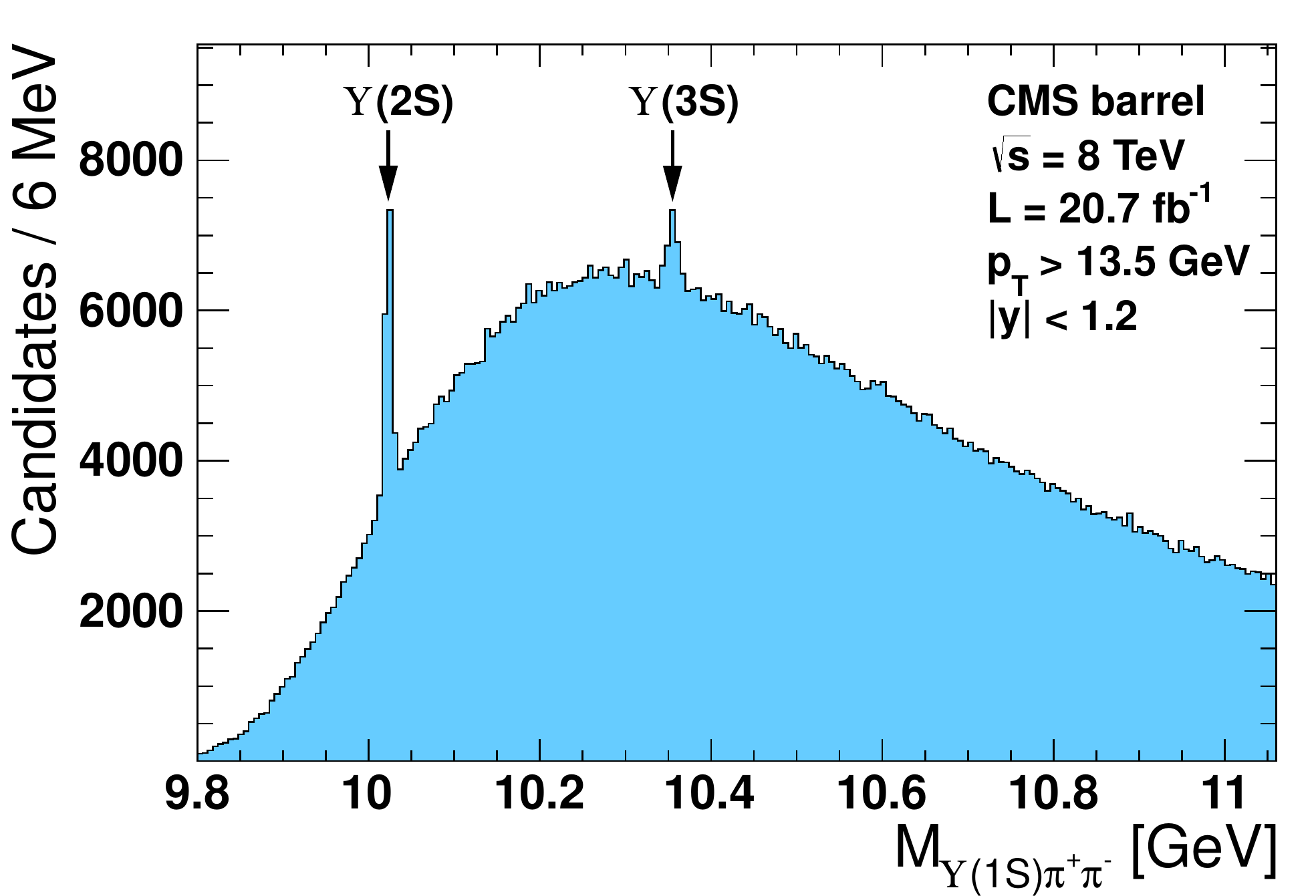}
    \includegraphics[width=0.49\textwidth]{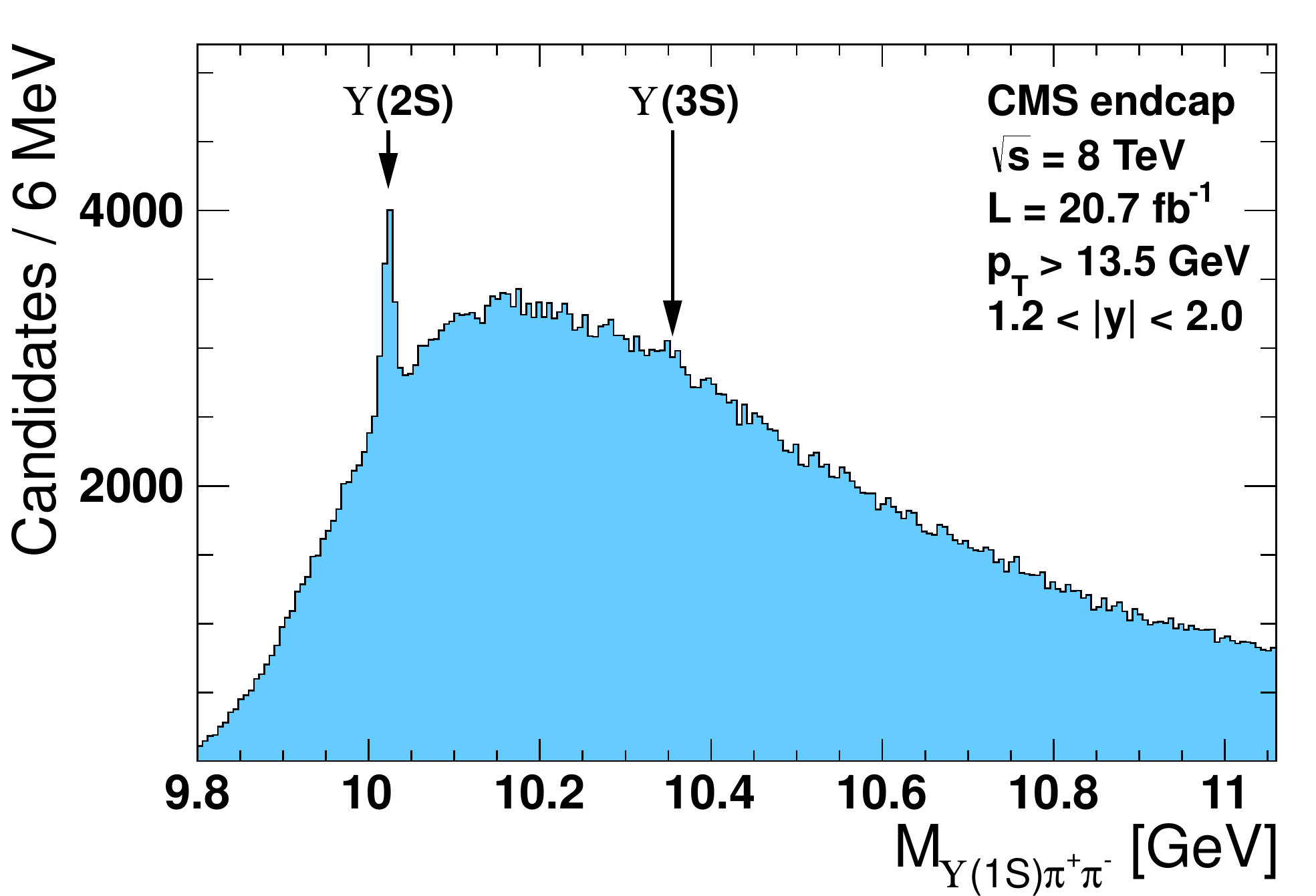}
    \caption{
The reconstructed invariant-mass distributions of the candidates
in the barrel (\cmsLeft) and endcap (\cmsRight) regions.
Peaks corresponding to $\PgUb\to\PgUa\Pgpp\Pgpm$
and $\PgUc\to\PgUa\Pgpp\Pgpm$ decays are indicated with the arrows.
    }
    \label{fig:mumupipi_mass}
\end{figure}

\begin{figure}[thb]
  \begin{center}
    \includegraphics[width=0.49\textwidth]{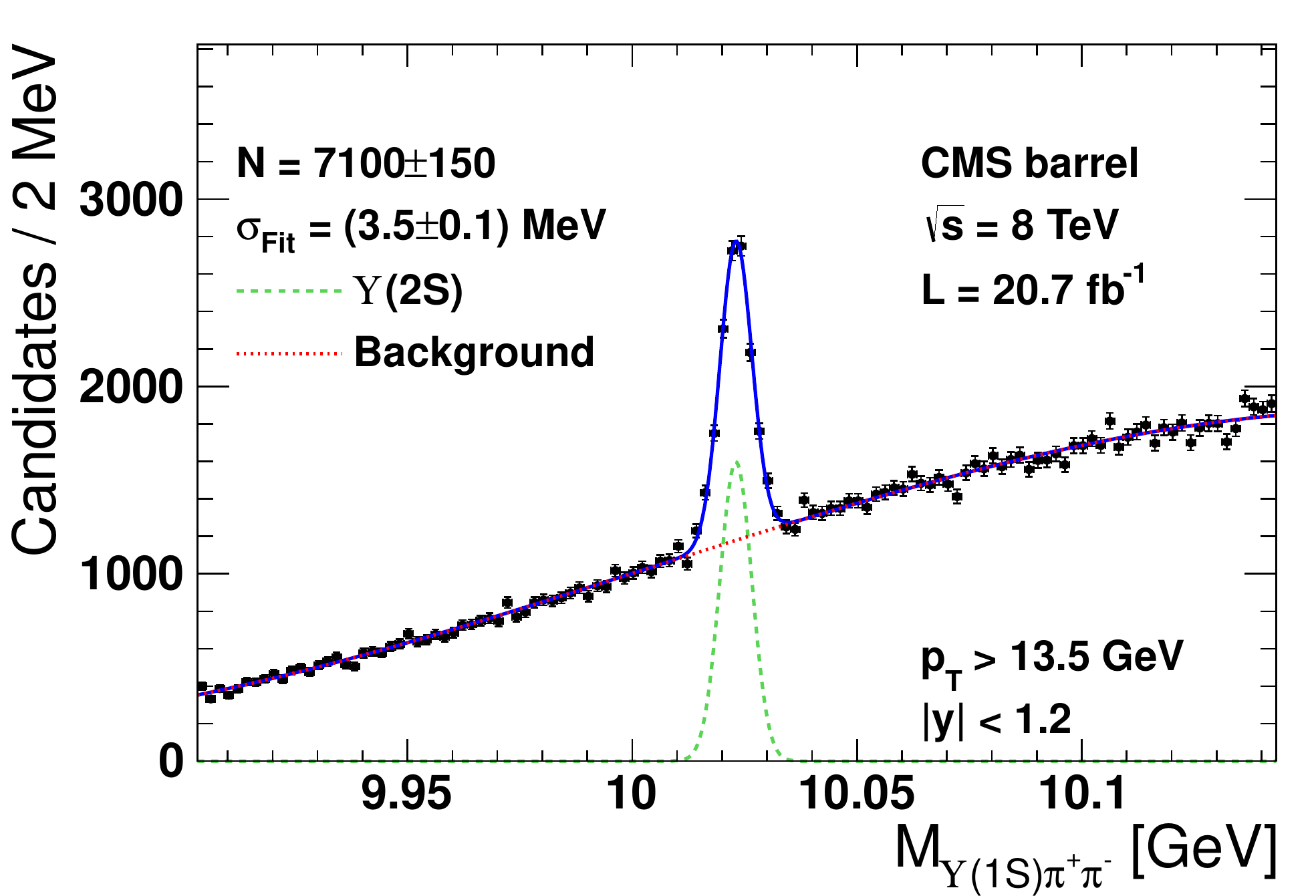}
    \includegraphics[width=0.49\textwidth]{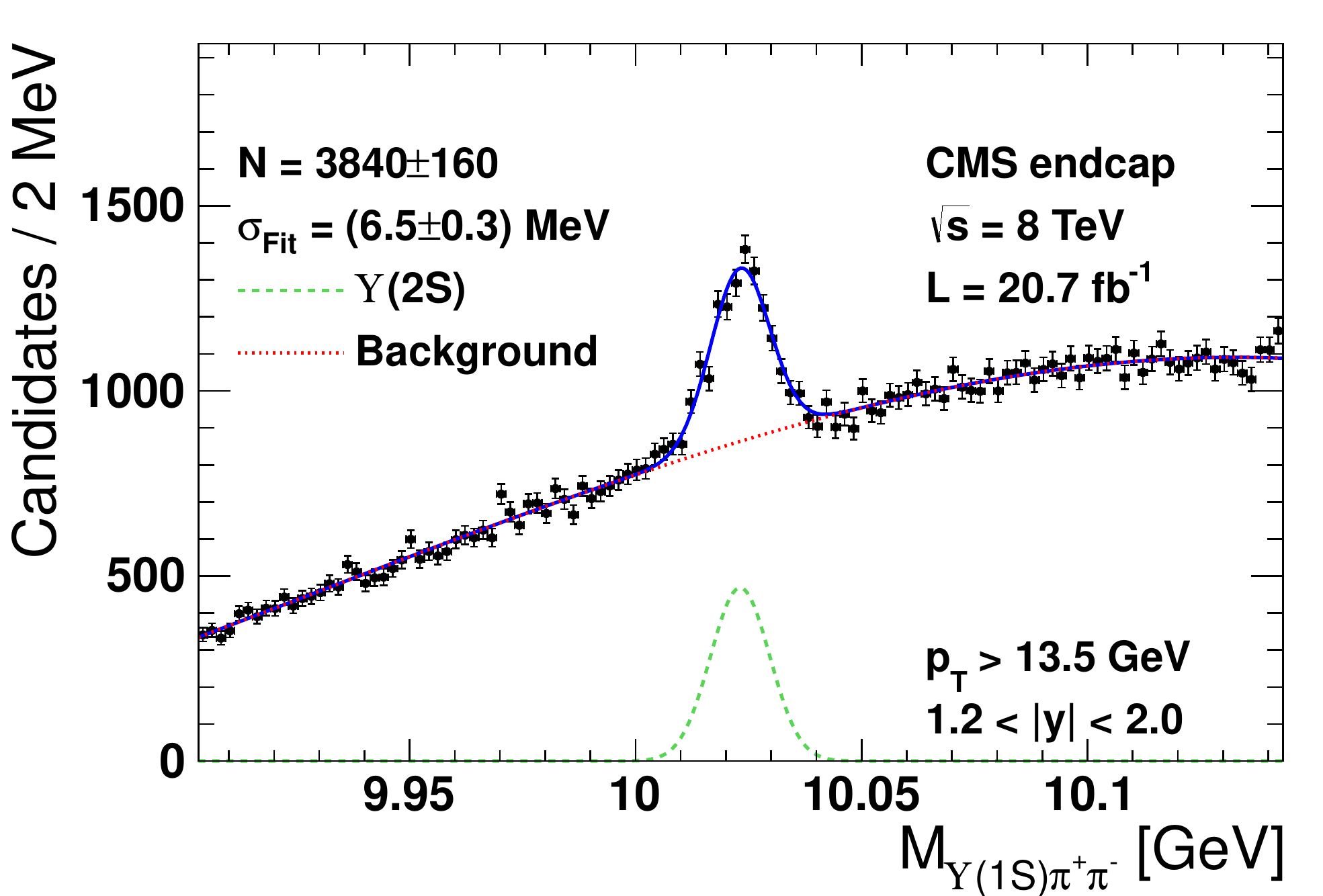}
    \caption{
The invariant-mass distributions of the candidates around the $\PgUb$ resonance for the barrel (\cmsLeft) and endcap (\cmsRight) regions.
The result from the fit is shown as a solid
curve; the $\PgUb$ and background contributions from the fit are shown separately as the
dashed and dotted curves, respectively.
}
    \label{fig:fit_upsilon2s}
  \end{center}
\end{figure}

Significant mass resolution and background level differences are observed for the $\PgUa\Pgpp\Pgpm$ candidates in the barrel ($\abs{y} < 1.2$) and endcap ($1.2 < \abs{y} < 2.0$) regions.
Therefore, the events are separated into these two classes. Figure~\ref{fig:mumupipi_mass}
shows the reconstructed invariant-mass distributions of the $\PgUa\Pgpp\Pgpm$ candidates in the barrel and endcap regions.
Apart from the peaks corresponding to the $\PgUb\to\PgUa\Pgpp\Pgpm$
and $\PgUc\to\PgUa\Pgpp\Pgpm$ decays,
the mass spectrum does not show any other outstanding structure.
Unbinned maximum-likelihood fits are
performed on the invariant-mass distributions
using single Gaussian functions to
describe the $\PgUb$ and $\PgUc$ states, and third-degree
polynomials to describe the combinatorial
background, whose parameters were left
free in the fit. The means and widths of the Gaussian
functions are allowed to float in the fit.
The resulting fit values of the widths are consistent with the mass resolutions
obtained from the simulated events.
The centers of the Gaussian functions are consistent with the world-average
values~\cite{Beringer:1900zz} for the two $\PgU$ masses.
The resulting numbers of $\PgUb\to\PgUa\Pgpp\Pgpm$
events are $7100\pm150$ and $3840\pm160$
for the barrel and endcap regions, where the uncertainties are statistical.
The invariant-mass distributions around the $\PgUb$ resonance
are shown in Fig.~\ref{fig:fit_upsilon2s} for the barrel
and endcap regions, with the results of the fits superimposed.

\section{Results}

The search for the $\cPXb$ is performed in the mass regions
10.06--10.31 and 10.40--10.99\GeV,
excluding the
mass intervals around the $\PgUb$ and $\PgUc$ resonances.
Fits are performed to the data by shifting the mean mass of the hypothetical $\cPXb$ signal in 10\MeV intervals and
letting the signal strength float.
The invariant-mass distribution of the reconstructed $\PgUa\Pgpp\Pgpm$ is modeled with a Gaussian function.
The intrinsic width of the $\cPXb$ is assumed to be small compared to
the detector mass resolution.  In the fits, the width of the Gaussian
is fixed to the values obtained from simulation.
Depending on the $\cPXb$ mass, the signal width is estimated to be in the range 3.8--10.6\MeV (6.8--16.4\MeV)
for the barrel (endcap) region.
The background distribution is modeled separately for the low-mass (10.06--10.31\GeV) and high-mass (10.40--10.99\GeV) regions with a third-degree
polynomial, whose coefficients are allowed to vary in the fit.
The signal yields and the coefficients of the
polynomials are determined from unbinned
maximum-likelihood fits to the
invariant-mass distributions for the barrel
and endcap regions.

For a given $\cPXb$ mass point, the
relationship between the $\cPXb$ and $\PgUb$ yields is given by
\begin{equation}\label{equ:master}
N_{\cPXb}^\text{obs} = R \times N_{\PgUb}^\text{obs} \times \frac{\epsilon_{\cPXb}}{\epsilon_{\PgUb}},
\end{equation}
where $N_{\cPXb}^\text{obs}$ and $N_{\PgUb}^\text{obs}$ are the observed $\cPXb$ and $\PgUb$ yields, respectively, and
$\epsilon_{\cPXb}/\epsilon_{\PgUb}$ is the ratio of overall efficiencies for $\cPXb$ and $\PgUb$ events.
This ratio uses the acceptance and the trigger and reconstruction efficiencies
estimated from simulated samples.
In the barrel region, the ratio of efficiencies increases from about 1 to 2
for a hypothetical $\cPXb$ mass in the range 10.06--10.31\GeV, mainly because of the increased acceptance
for higher masses, and remains around 2 in the range 10.40--10.99\GeV.
In the endcap region, the ratio of efficiencies stays around 1 for all the $\cPXb$ mass values considered.

In tests of statistical significance,
the p-value is the probability of obtaining a signal strength as large as (or larger than)
the one that was actually observed, assuming that there is no signal.
 A signal-like distribution will result in a low observed p-value.
In this analysis, the p-value is evaluated from simultaneous
signal-plus-background fits to the observed invariant-mass distributions
in the barrel and endcap regions.
Significances of the $\cPXb$ signal are evaluated for each hypothetical $\cPXb$ mass.
Given no strong hint of a signal in the present data, an upper limit on $R$, the ratio of the production cross sections times
branching fractions of the $\cPXb$ and $\PgUb$, is calculated.

\subsection{Systematic uncertainties}\label{sec:systematics}

Several sources of systematic uncertainties are considered.
The major sources are from the modeling of the signal decay,
which includes the dipion invariant-mass distribution
and the $\cPXb$ mass resolution, the signal polarization, and the background shape.

The dipion distributions in the simulated $\PgUb$ and $\cPXb$ samples are reweighted according to the $\PgUb$ data from CLEO~\cite{CroninHennessy:2007zz}, but the actual distribution of the $\cPXb$ is unknown.
This affects the value of the
efficiency ratio $\epsilon_{\cPXb}/\epsilon_{\PgUb}$ in Eq.~\ref{equ:master}.
Several alternative models have been implemented, including a $\PgUa\rho$ model, a model using the dipion invariant-mass distribution measured in $\cPXn$ decay~\cite{Chatrchyan:2013cld}, and a three-body $S$-wave model.
Since the actual quantum numbers for the $\cPXb$ state are not known, 
the dipion invariant-mass distribution in the $\cPXb \to \PgUa\rho$ decay is used in the systematic studies, 
which is similar to the $\cPXn \to \JPsi\rho$ decay.
The $\cPXb \to \PgUa\rho$ process is modeled with a uniform two-body phase-space decay.
The dipion mass distribution from $\cPXn$ decay is scaled according to the mass difference between the $\cPXb$
and the $\PgUa$.
A comparison between the alternative models and the default model using the $\PgUb$ distribution
leads to differences in the $\epsilon_{\cPXb}/\epsilon_{\PgUb}$ efficiency ratio
of up to 20\% depending on the $\cPXb$ mass, which is included as a systematic uncertainty.
The reconstruction efficiency as a function of $\cPXb$ mass is modeled with a simple analytical function.
The systematic uncertainty in this modeling is estimated by comparing two different functions and is found to be negligible.

The $\PgUb$ mass resolutions determined in data and simulation are consistent with each other.
The statistical uncertainty in the $\PgUb$ mass resolution from data of 2.9\% (4.6\%) in
the barrel (endcap) region is larger than the difference between the measured and simulated values.
The statistical uncertainty is taken as the systematic uncertainty from this source.
While a single Gaussian function is used in the default modeling of the signal, a sum of two Gaussians is used as an alternative model,
and the differences between the respective fits are taken as systematic uncertainties.

A recent CMS measurement~\cite{Chatrchyan:2012woa} shows that $\PgUb$ mesons are produced
with negligible polarization.
The daughter $\PgUa$ mesons are expected to have a similar polarization~\cite{Faccioli:2012kp}.
However, the expected polarization of the $\cPXb$ is unknown.
Signal efficiencies evaluated using
a simulated sample generated
with unpolarized $\cPXb$ are
compared with efficiencies for the extreme cases of full transverse
and full longitudinal polarizations in the helicity frame, assuming that the polarization of the daughter
$\PgUa$ is the same as that of the mother $\cPXb$.
The largest efficiency difference of 25\%
is taken as the systematic uncertainty from this source.

The fit model is composed of a background component, with
floating coefficients, and a signal model, with the signal strength as a free parameter.
The uncertainties in the coefficients from the fits are included as a systematic uncertainty
in the statistical analysis. Furthermore,
an alternative background parameterization, determined from
a background-only fit to
the candidates reconstructed with same-sign pions ($\PgUa\Pgpp\Pgpp$ and $\PgUa\Pgpm\Pgpm$) is also considered.
The difference between the default and alternative background parameterizations is included as one of the systematic uncertainties.

Other systematic uncertainties,
such as the uncertainty caused by the dependence
of the efficiencies on the number of pp interactions
per event (with an average of ${\approx}21$ interactions), have been considered and found to be negligible.
Systematic uncertainties in the acceptance and trigger efficiency largely cancel out in the ratio $R$.
As a check, the $\PgUb$ yields, normalized to the integrated luminosity, are found to be stable for the different data-taking periods.

\subsection{Determination of p-values and upper limits}\label{sec:limit}

\begin{figure}[thp]
  \begin{center}
    \includegraphics[width=\cmsFigWidth]{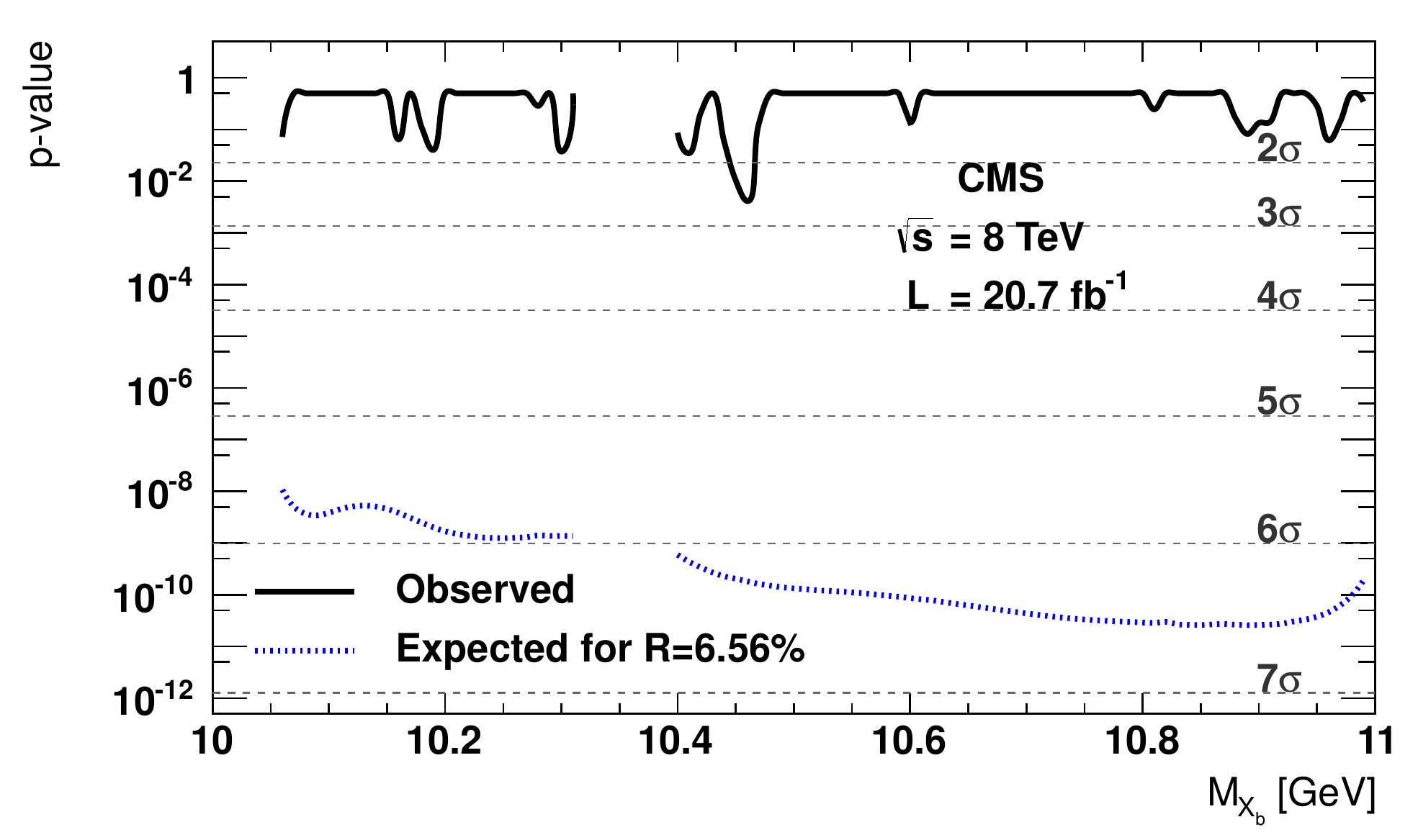}
    \caption{Observed (solid curve) and expected for $R=6.56\%$ (dotted curve) local p-values, as a function of the assumed $\cPXb$ mass.}
    \label{fig:p-values}
  \end{center}
\end{figure}

\begin{figure}[thp]
  \begin{center}
    \includegraphics[width=\cmsFigWidth]{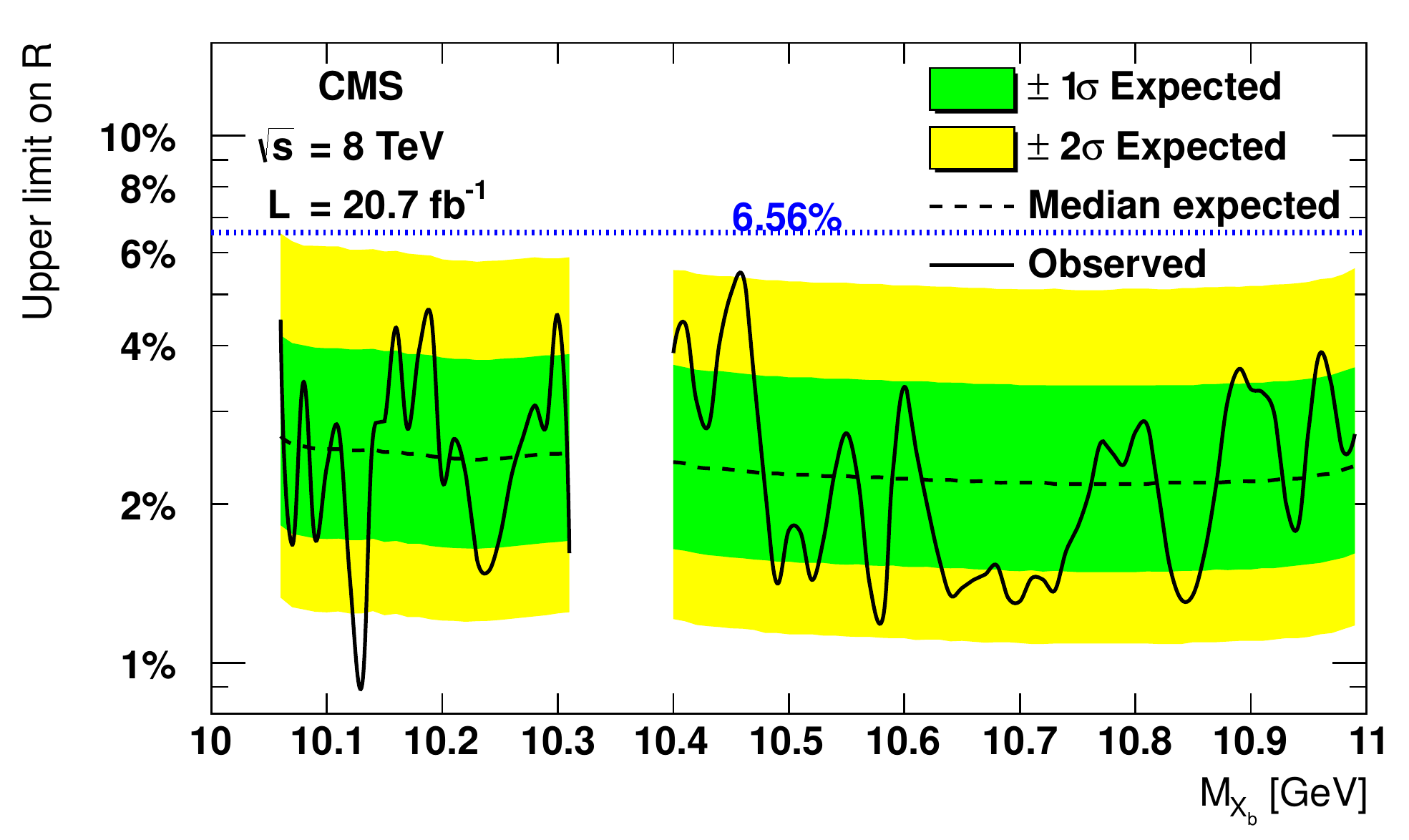}
    \caption{
    Upper limits at the 95\% confidence level on $R$,
the production cross section for the
$\cPXb$ times its branching fraction to $\PgUa\Pgpp\Pgpm$,
relative to the $\PgUb$, as a function of the
$\cPXb$ mass.
    The solid curve shows the observed limits,
    while the dashed curve represents
    the expected limits in the absence of a signal, with the two
shaded regions giving the $\pm1$ and $\pm2$
standard deviation uncertainties on the expected
limits. The measured value for the analogous $\cPXn$ to $\psi(2S)$ ratio of 6.56\% is
shown by the dotted line.}
    \label{fig:limit}
  \end{center}
\end{figure}

The local p-values are calculated using an asymptotic approach~\cite{Cowan:2010js} with the signal and background
models described above and combining the results of the fits to the barrel and endcap regions.
The systematic uncertainties mentioned above are implemented as nuisance parameters in the fit, assuming
log-normal or flat priors.
The expected discovery potential is
estimated by injecting various amounts of
signal events into the fits and
evaluating the resulting p-values.
The expected signal significance for the
assumption $R=6.56\%$, motivated by the
ratio of production cross sections times branching fractions for $\cPXn$
and $\psi(2S)$ reported in Ref.~\cite{Chatrchyan:2013cld},
is larger than
five standard deviations ($\sigma$) across the
explored $\cPXb$ mass range, as shown by the dashed
curve in Fig.~\ref{fig:p-values}.
The observed p-values displayed
in Fig.~\ref{fig:p-values} by the solid line show no indication of an $\cPXb$ signal.
The smallest local p-value is 0.004 at 10.46\GeV, corresponding to a statistical significance of $2.6\sigma$,
which is reduced to $0.8\sigma$ when taking into account the ``look-elsewhere effect"~\cite{Gross:2010qma}.
The expected and observed 95\% confidence level upper limits on $R$,
derived using a modified frequentist approach
($\mathrm{CL}_{S}$)~\cite{cls, Junk:1999kv},
are shown in Fig.~\ref{fig:limit} as a function of the $\cPXb$ mass.
The observed upper limits on $R$ are in the range 0.9--5.4\% at 95\% confidence level.
The expected upper limits, which are derived for a pure background hypothesis, are less stringent than
those obtained from the p-value calculations.
This is because the p-value calculations are only concerned with the probability of the
background fluctuating to a signal-like peak in the invariant-mass distribution,
while the upper limits on $R$ also
include the systematic uncertainties in the signal normalization from
the signal decay model and $\cPXb$ polarization assumptions.

\section{Summary}

A search for an exotic bottomonium state in the decay channel $\cPXb \to \PgUa\Pgpp\Pgpm$, followed by
$\PgUa \to \Pgmp\Pgmm$, in pp collisions at $\sqrt{s} = 8$\TeV at the LHC has been presented. This analysis was performed using data collected by the CMS experiment, corresponding to an integrated luminosity of 20.7\fbinv. Candidates were
reconstructed from two identified muons and two
additional charged tracks assumed to be pions.
The search was conducted in the
kinematic region $\pt(\PgUa\Pgpp\Pgpm)>13.5$\GeV and $\abs{y(\PgUa\Pgpp\Pgpm)}<2.0$.
The $\PgUb\to \PgUa\Pgpp\Pgpm$ process was used as a normalization channel,
canceling many of the systematic uncertainties. Excluding the
known $\PgUb$ and $\PgUc$ resonances, no significant excess above the background
was observed for $\cPXb$ masses between
10 and 11\GeV. The expected sensitivity of the analysis
was greater than five standard deviations for the explored $\cPXb$ mass range, if the
relative signal strength is comparable to the corresponding
value for the $\cPXn$ of 6.56\%.
The resulting 95\% confidence level upper limit on the ratio
$\sigma(\Pp\Pp \to \cPXb \to \PgUa\Pgpp\Pgpm)/\sigma(\Pp\Pp \to \PgUb \to \PgUa\Pgpp\Pgpm)$
is in the range 0.9--5.4\%, depending on the assumed $\cPXb$ mass.
These are the first upper limits on the production of
a possible $\cPXb$ at a hadron collider.

\section*{Acknowledgements}

We congratulate our colleagues in the CERN accelerator departments for the excellent performance of the LHC and thank the technical and administrative staffs at CERN and at other CMS institutes for their contributions to the success of the CMS effort. In addition, we gratefully acknowledge the computing centres and personnel of the Worldwide LHC Computing Grid for delivering so effectively the computing infrastructure essential to our analyses. Finally, we acknowledge the enduring support for the construction and operation of the LHC and the CMS detector provided by the following funding agencies: BMWF and FWF (Austria); FNRS and FWO (Belgium); CNPq, CAPES, FAPERJ, and FAPESP (Brazil); MEYS (Bulgaria); CERN; CAS, MoST, and NSFC (China); COLCIENCIAS (Colombia); MSES (Croatia); RPF (Cyprus); MoER, SF0690030s09 and ERDF (Estonia); Academy of Finland, MEC, and HIP (Finland); CEA and CNRS/IN2P3 (France); BMBF, DFG, and HGF (Germany); GSRT (Greece); OTKA and NKTH (Hungary); DAE and DST (India); IPM (Iran); SFI (Ireland); INFN (Italy); NRF and WCU (Republic of Korea); LAS (Lithuania); CINVESTAV, CONACYT, SEP, and UASLP-FAI (Mexico); MSI (New Zealand); PAEC (Pakistan); MSHE and NSC (Poland); FCT (Portugal); JINR (Armenia, Belarus, Georgia, Ukraine, Uzbekistan); MON, RosAtom, RAS and RFBR (Russia); MSTD (Serbia); SEIDI and CPAN (Spain); Swiss Funding Agencies (Switzerland); NSC (Taipei); ThEPCenter, IPST and NSTDA (Thailand); TUBITAK and TAEK (Turkey); NASU (Ukraine); STFC (United Kingdom); DOE and NSF (USA).

Individuals have received support from the Marie-Curie programme and the European Research Council and EPLANET (European Union); the Leventis Foundation; the A. P. Sloan Foundation; the Alexander von Humboldt Foundation; the Belgian Federal Science Policy Office; the Fonds pour la Formation \`a la Recherche dans l'Industrie et dans l'Agriculture (FRIA-Belgium); the Agentschap voor Innovatie door Wetenschap en Technologie (IWT-Belgium); the Ministry of Education, Youth and Sports (MEYS) of Czech Republic; the Council of Science and Industrial Research, India; the Compagnia di San Paolo (Torino); the HOMING PLUS programme of Foundation for Polish Science, cofinanced by EU, Regional Development Fund; and the Thalis and Aristeia programmes cofinanced by EU-ESF and the Greek NSRF.

\bibliography{auto_generated}   

\cleardoublepage \appendix\section{The CMS Collaboration \label{app:collab}}\begin{sloppypar}\hyphenpenalty=5000\widowpenalty=500\clubpenalty=5000\textbf{Yerevan Physics Institute,  Yerevan,  Armenia}\\*[0pt]
S.~Chatrchyan, V.~Khachatryan, A.M.~Sirunyan, A.~Tumasyan
\vskip\cmsinstskip
\textbf{Institut f\"{u}r Hochenergiephysik der OeAW,  Wien,  Austria}\\*[0pt]
W.~Adam, T.~Bergauer, M.~Dragicevic, J.~Er\"{o}, C.~Fabjan\cmsAuthorMark{1}, M.~Friedl, R.~Fr\"{u}hwirth\cmsAuthorMark{1}, V.M.~Ghete, N.~H\"{o}rmann, J.~Hrubec, M.~Jeitler\cmsAuthorMark{1}, W.~Kiesenhofer, V.~Kn\"{u}nz, M.~Krammer\cmsAuthorMark{1}, I.~Kr\"{a}tschmer, D.~Liko, I.~Mikulec, D.~Rabady\cmsAuthorMark{2}, B.~Rahbaran, C.~Rohringer, H.~Rohringer, R.~Sch\"{o}fbeck, J.~Strauss, A.~Taurok, W.~Treberer-Treberspurg, W.~Waltenberger, C.-E.~Wulz\cmsAuthorMark{1}
% ...
\vskip\cmsinstskip
% ...
1:~~Also at Vienna University of Technology, Vienna, Austria\\
2:~~Also at CERN, European Organization for Nuclear Research, Geneva, Switzerland\\
% ...

\end{sloppypar}
\end{document}